\documentclass[manuscript]{acmart}

\AtBeginDocument{%
  \providecommand\BibTeX{{%
    \normalfont B\kern-0.5em{\scshape i\kern-0.25em b}\kern-0.8em\TeX}}}

\setcopyright{acmcopyright}
\copyrightyear{2024}
\acmYear{2024}
\acmDOI{XXXXXXX.XXXXXXX}

\begin{document}

\title[Generative Artificial Intelligence in Learning Analytics]{Generative Artificial Intelligence in Learning Analytics: Contextualising Opportunities and Challenges through the Learning Analytics Cycle}

\author{Lixiang Yan}
\affiliation{%
  \institution{Monash University}
  \country{Australia}
}

\author{Roberto Martinez-Maldonado}
\affiliation{%
  \institution{Monash University}
  \country{Australia}
}

\author{Dragan Gašević}
\affiliation{%
  \institution{Monash University}
  \country{Australia}
}

\renewcommand{\shortauthors}{Yan et al.}

\begin{abstract}

Generative artificial intelligence (GenAI), exemplified by ChatGPT, Midjourney, and other state-of-the-art large language models and diffusion models, holds significant potential for transforming education and enhancing human productivity. While the prevalence of GenAI in education has motivated numerous research initiatives, integrating these technologies within the learning analytics (LA) cycle and their implications for practical interventions remain underexplored. This paper delves into the prospective opportunities and challenges GenAI poses for advancing LA. We present a concise overview of the current GenAI landscape and contextualise its potential roles within Clow's generic framework of the LA cycle. We posit that GenAI can play pivotal roles in analysing unstructured data, generating synthetic learner data, enriching multimodal learner interactions, advancing interactive and explanatory analytics, and facilitating personalisation and adaptive interventions. As the lines blur between learners and GenAI tools, a renewed understanding of learners is needed. Future research can delve deep into frameworks and methodologies that advocate for human-AI collaboration. The LA community can play a pivotal role in capturing data about human and AI contributions and exploring how they can collaborate most effectively. As LA advances, it is essential to consider the pedagogical implications and broader socioeconomic impact of GenAI for ensuring an inclusive future.

\end{abstract}


\maketitle

\section{Introduction}

Generative artificial intelligence (GenAI), catalysed by the public release of ChatGPT in November 2022, has sparked a new era of innovation and productivity, particularly in the educational sector. Tools like ChatGPT, powered by state-of-the-art large language models (LLMs), have illustrated their potential in reshaping the educational landscape, fostering enhanced engagement and interactive learning experiences \cite{van2023chatgpt, kasneci2023chatgpt}. A comprehensive scoping review has identified 53 different applications of LLMs in automating educational tasks \cite{yan2023practical}. Subsequent studies delving into ChatGPT and other LLM tools illustrated their capabilities in delivering comprehensive feedback, surpassing the efficacy of human educators in articulating student performance \cite{dai2023can}. These models also outperformed average students in reflective writing tasks \cite{li2023can} and innovated conversational assessments that could potentially transform conventional digital formative assessments \cite{yildirim2023conversation}. Yet, the spotlight on ChatGPT represents just the tip of the iceberg in the vast GenAI ecosystem. The maturity of text-to-image GenAI models like DALL-E 2 and Midjourney has facilitated the creation of dynamic teaching resources, supporting multimedia learning in specialised fields such as medical education \cite{mazzoli2023enhancing} and craft training \cite{vartiainen2023using}. Meanwhile, speech-to-text models, such as OpenAI's Whisper, could potentially transform educational content transcription \cite{rao2023transcribing} and automate the capture of collaborative discourse \cite{cao2023comparative}. The GenAI landscape further broadens with text-to-code models like OpenAI's Codex and text-to-audio models like Meta's Voicebox \cite{le2023voicebox}, offering learning analytics (LA) researchers and practitioners a diverse toolkit for data analytics and communication. As these tools continue to evolve, their integration into the LA cycle \cite{clow2012learning} holds promise in addressing some of the pressing challenges in the field, especially the lack of attempts to intervene in the learning environment \cite{motz2023lak}. Nevertheless, how these novel technologies can be embedded in the LA cycle \cite{clow2012learning} and benefit the development of practical LA solutions remains largely unknown. This position paper explores the opportunities and challenges GenAI poses for advancing learning analytics. The contribution of this work to the LA community is that we contextualised the opportunities and challenges of GenAI within Clow's \cite{clow2012learning} LA cycle and summarised six endeavours for future research. These contributions could pave the way for future LA research and practice to integrate GenAI tools effectively, ensuring that they align with educational goals and uphold the principles of inclusive education while navigating the ethical complexities introduced by these advanced technologies.
\section{Current Landscape of Generative AI}

We first provide a concise overview of the current landscape of GenAI, focusing on the state-of-the-art models, promising applications, and supporting infrastructures instead of the underlying algorithms or technological mechanisms, as the aim of this paper is to explore the practical implications of GenAI in LA. While LLMs like GPT-4 and Llama-2 have attracted most of the attention in academic research, especially in education \cite{van2023chatgpt, kasneci2023chatgpt, choi2023chatgpt}, GenAI's capabilities extend beyond merely text-to-text generation but also to other modalities, such as image, video, and speech generation \cite{chiu2023impact, mazzoli2023enhancing, rao2023transcribing}. Within text-based GenAI, there are many application scenarios beyond chatbots like ChatGPT. For example, the idea and exploration into AI agents (e.g., AgentGPT \cite{rework2023agentgpt}) were based on GPT-4. These AI agents are autonomous and adaptive AI that work towards a goal and make decisions independently without continuous user input. They could potentially self-generate prompts based on the outputs of the previous prompt, making them a potential candidate for working proactively with humans \cite{hauptman2023adapt} and simulating human behaviours \cite{park2023generative}. Similar to text generation, code generation also utilises LLMs but with additional training on code-related content (e.g., OpenAI's Codex and Meta's Code Llama). These text-to-code generation models excel when paired with a code interpreter like in ChatGPT-4, where they can automatically conduct data analysis on provided datasets and with natural language prompts \cite{wang2023code}. 

On the other hand, image generation models like Midjourney, Stable Diffusion, and Dall-E are capable of generating high-quality images from natural language text prompts. While these text-to-image generation models have been previously criticised for their inconsistency and lack of controllability over multiple generations, which might limit the practical utility in creating narrated or specialised contents (e.g., illustration of cardiac arrest training) \cite{mazzoli2023enhancing, chiu2023impact}, recent progress, such as Midjourney's new 'inpainting' function and the latest Dall-E 3 have made them more controllable and consistent for image generation tasks. Likewise, video generation models like Runway's Gen-2 and Meta's Make-A-Video \cite{singer2022make} are capable of creating video from text, image, or a combination of both. However, the practical utility of this line of GenAI models still needs to be discovered as the technologies become more mature and accessible. Speech-wise, text-to-speech models like Meta's Voicebox \cite{le2023voicebox} are capable of generating and synthesising audio that mirrors the tone and voice of a human in multiple different languages. Whereas, speech-to-text models like OpenAI's Whisper \cite{gris2023evaluating} have demonstrated their ability to automatically generate transcripts from audio inputs with performance comparable and even surpass human transcription when the audio is severally influenced by background noise \cite{gris2023evaluating}. Preliminary studies that utilised Whisper to automatically transcribe educational videos have already shown promising results in enriching the modality of existing educational resources \cite{rao2023transcribing}.

Apart from these foundational generative models, there are also many supporting infrastructures that have been deployed to address some of their challenges. In particular, for LLMs, the emergence of vector embedding databases like Milvus (milvus.io) aims to address the limited context length of most existing models (e.g., GPT-3.5/4 and Llama 2). This limitation is often imposed on these LLMs to optimise computational resources as the self-attention mechanism, which is core to Transformer models, struggles with very long sequences due to its quadratic complexity \cite{vaswani2017attention}. Vector embedding is the representation of objects, such as words or entities, as vectors in a continuous vector space to capture semantic or relational meaning \cite{mikolov2013distributed}. Such embedding databases can also act as an external knowledge database for LLMs, where they can draw knowledge from and use these databases to limit the context of AI responses. For example, a common workflow for developing in-document questions and response systems involves using an application framework (e.g., Langchain and Llama Index) to segment the document content into different chucks, generate the vector embedding of each chuck, and store them in an embedding database, which can be retrieved and compared with the embedding of prompts to identify the most relevant chunks. Together, the prompt and these identified chucks can be sent to LLMs like GPT-4 to generate contextually accurate and reliable responses \cite{cui2023chatlaw}.

\section{Generative AI in Learning Analytics}

\begin{figure}[htbp]
    \centering
      \includegraphics[width=.65\textwidth]{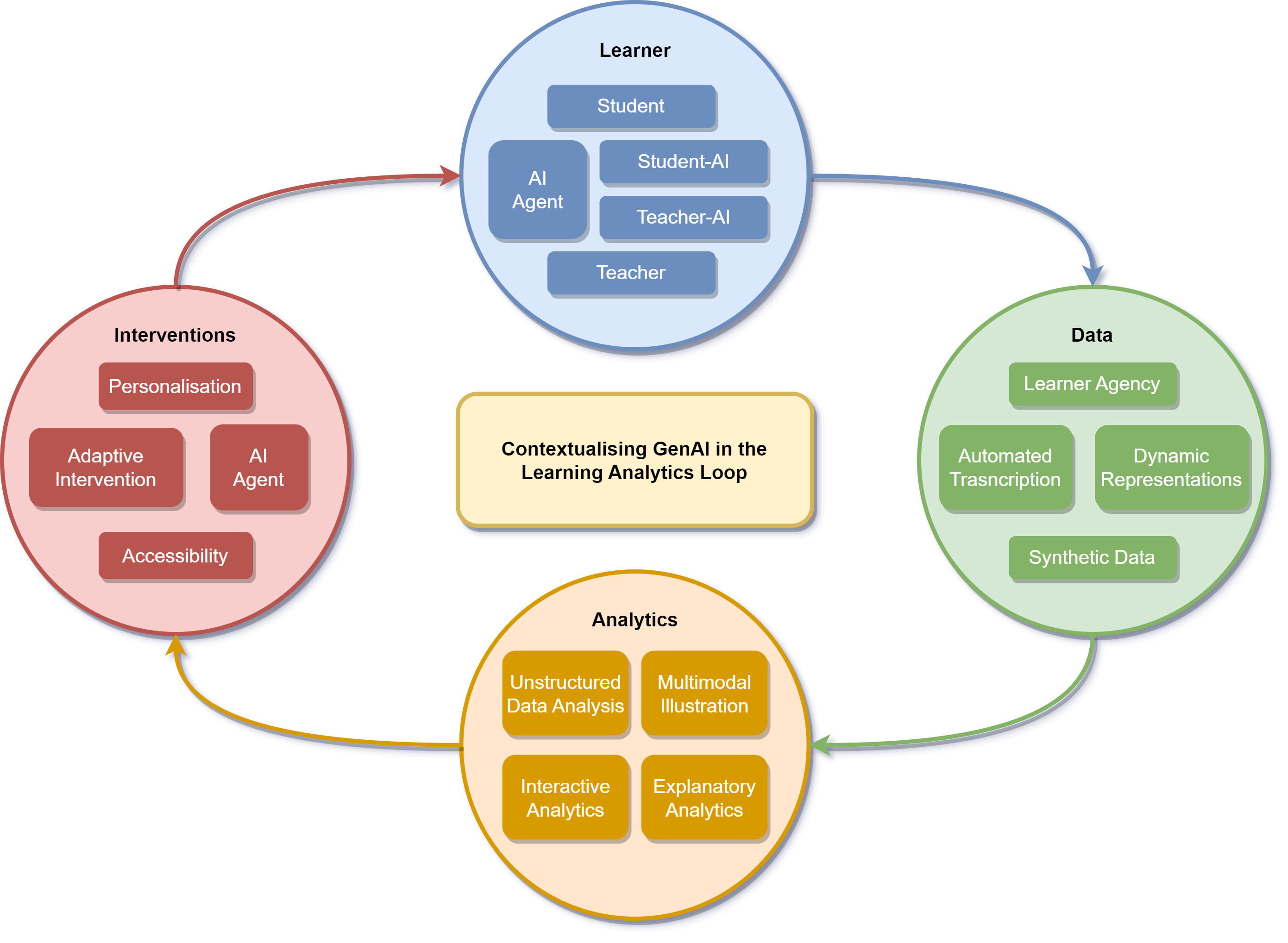}
    \caption{Contextualising the implications of generative artificial intelligence within each step of Clow's learning analytics cycle.}
    \label{fig:LAcycle}
\end{figure}

Clow's \cite{clow2012learning} generic LA cycle provides a well-grounded framework for exploring the potential influence of GenAI in LA research and practices as it highlights the four essential steps in LA, including identifying the learner, collecting the relevant data, generating and presenting analytics, and delivering and evaluating interventions. As shown in figure \ref{fig:LAcycle}, GenAI have multiple implications for each of these steps. The following sections elaborated on each of these steps and discussed the relevant theoretical and empirical basis.

\vspace{-10pt}

\subsection{Learner}

The foundation of LA is centred on understanding learning and learners \cite{gavsevic2015let}. A pivotal initial step in the LA cycle is identifying the learner and the contexts in which learning occurs. This step is essential for ensuring that LA aligns with its primary objective: to enhance learning for learners in learning environments \cite{lang_what_2022}. Before the prevalence of GenAI, pinpointing the learner was relatively straightforward. Typically, a learner could be an individual pursuing a degree or education at a university or in school, or participating in a Massive Open Online Course (MOOC), other online education platforms, or even physical space. Under such circumstances, LA researchers and practitioners could confidently ascertain the learner's identity and their learning environment. Given this clarity, the data collected, either from the learning products (e.g., assessments) or the learning processes (e.g., interactions in online forums), can be deemed reliable to a certain degree for modelling learners' metacognitive (e.g., self-regulated learning \cite{roll2015understanding, winne_learning_2022} and socially shared regulation of learning \cite{jarvela2015enhancing, jarvela2023predicting}), cognitive (e.g., memory and reflection \cite{kovanovic2018understand, gibson2017reflective}), and affective processes (e.g., emotion and motivation \cite{suero2014emotion, dmello_emotional_2022}). 

As GenAI becomes more capable, the underlying assumption of knowing learners' identity might face challenges. GenAI tools like ChatGPT and Claude, which students can utilise for various learning tasks, have blurred the boundary between learners and tools. These tools have been demonstrated to generate learning products indicative of high metacognitive, cognitive, and social skills. Specifically, recent research indicates that ChatGPT can produce reflective writings that surpass the quality produced by average undergraduate and graduate pharmacy students. Notably, these AI-generated writings were indistinguishable from student submissions by both educators and state-of-the-art AI detection systems (e.g., OpenAI's AI text classier) \cite{li2023can}. Such capabilities also extend to structured and high-stake assessments. Preliminary findings suggest that GenAI tools, like GPT-3 and ChatGPT, successfully passed the United States Medical Licensing Exam \cite{kung2023performance}, passed four real exams at the University of Minnesota Law School \cite{choi2023chatgpt}, and answered basic operations management and process-analysis questions from Wharton School of Business examinations \cite{terwiesch2023would}. While these test-taking performances might be expected considering potential data contamination, where the vast training data of these LLMs already contained the answers, there is also evidence showing signs of emotional intelligence in GenAI. A study highlighted that AI-generated responses to patient queries in an online forum were deemed more empathetic and of higher quality than responses from physicians by experts \cite{ayers2023comparing}. 

Given GenAI's capabilities, the products of learning, such as written assignments and test scores, can easily be influenced. On the other hand, the processes of learning, which include the steps, strategies, and cognitive activities learners engage in, may remain more resistant to GenAI interference. While GenAI can provide learners with information or even automate certain tasks, the actual cognitive processes, emotions, and motivations that constitute their learning process are still human-centric \cite{kasneci2023chatgpt, jarvela2023human}. However, as GenAI become more sophisticated, there is also the potential for it to not only complete learning products but also mimic certain learning processes. For instance, a student might employ an AI tool to read and summarise content, bypassing the cognitive process of comprehension. While this might lead to higher task efficiency, it raises questions about the depth and authenticity of their learning process and challenges the very essence of what it means to be a 'learner' \cite{rudolph2023chatgpt}.

Therefore, instead of solely focusing on validating learners' identities, the LA community should pivot towards understanding the evolving nature of the learner in the age of GenAI. Learners will increasingly be expected to interact with and utilise GenAI tools in their professional capacities \cite{kasneci2023chatgpt}. This calls for a shift in our conceptualisation of learners. Drawing from novel insights and frameworks like hybrid human-AI regulation of learning \cite{molenaar2022concept}, artificial and human cognition \cite{siemens2022human}, human and AI collaboration in socially shared regulation of learning \cite{jarvela2023human}, hybrid intelligence \cite{jarvela2023hybrid}, and human-AI teaming \cite{zhang2021ideal}, we advocate for a renewed understanding of learners. Rather than seeing GenAI as a threat to learning integrity, we should view it as an opportunity to transform learning in the context of human-AI collaboration \cite{joksimovic2023opportunities, national2021human}. This evolving learner definition has significant implications for other stages of the LA cycle, offering exciting avenues for both theoretical and empirical research. The LA community can play a pivotal role in capturing data about human and AI contributions and exploring how they can collaborate most effectively.

\subsection{Data}

LA is a data-centred field that leverages learners' data to understand and optimise their learning process \cite{lang_what_2022}. Yet, not all data is relevant. The second step of the LA cycle is to capture relevant data for understanding and improving learners' learning process \cite{clow2012learning}. These data typically involve survey data (e.g., self-report measures of learner perceptions and self-efficacy), trace data (e.g., digital traces in online learning platforms like clicks and page views), and text data (e.g., forum posts, discourse in intelligence tutoring system, and written assessments) \cite{gray_practitioners_2022}. Beyond these conventional data streams, recent advancements in multimodal LA and game LA also highlighted the educational value contained in learners' physical and physiological traces (e.g., gesture, speech, and heart rate) \cite{ochoa_multimodal_2022}, and game progression, achievements, and decisions \cite{reardon_game_2022}. In terms of capturing relevant learner data, while GenAI could potentially introduce complications and novel challenges, such as attributing learner agency, it can also open up new opportunities for transforming and augmenting existing data-capturing processes in LA, including formatting unstructured text data, expanding the modality of learner data, and support the generating of synthetic data for model training and open-science initiatives. The following sections elaborate on each of these challenges and opportunities.

\subsubsection{Learner Agency}

As the boundaries between AI and human learners blur, the significance of capturing data about the learning processes and accounting for learner agency in LA becomes even more pronounced \cite{jaaskela2021student}. Learner agency is defined as learners' capacity to act independently and make decisions regarding their learning \cite{bandura1989human}. For LA to optimise learning for human learners, it is essential to understand their agency during the learning process as it determines what they choose to learn and how they conduct learning, navigate challenges, and integrate feedback \cite{reeve2013students}. GenAI poses a significant challenge in identifying when and how a learner exerts agency over their learning process as learners may offload or collaborate with GenAI to conduct various learning tasks \cite{rudolph2023chatgpt}. For example, some students may use GenAI tools to improve the grammar and presentation of their reflective writing, while others may ask these tools to generate the entire reflective assessment, undermining the educational purpose of such tasks. LA should ideally be able to capture and distinguish such differences in learner agency as these differences also contain significant insights. For example, based on the social cognitive perspective of self-regulation \cite{zimmerman2000attaining}, a learner's decision to use GenAI tools for specific tasks but not others could potentially offer insights into their self-efficacy, values, and strategic decision-making processes. Likewise, from established work in goal orientation and self-regulated learning (e.g., traced goal orientation \cite{zhou2012modeling}), learners' interactions with these GenAI tools can also provide some insights into their metacognitive processes, such as when they seek GenAI for assistance, when they disagree, challenge, or question AI-generated results, and how they integrate AI-generated content with their unique understanding and perspectives. As GenAI becomes increasingly prevalent, the failure to capture learner agency could lead to biased and unfair interpretations of learning data and conceal the rich interplay of decision-making, motivation, and self-regulation \cite{turner2008does}. Given this context, the LA community must urgently build upon the established frameworks (e.g., Winne and Hadwin's COPES model \cite{winne1998studying}) and emerging research on human-AI regulation \cite{molenaar2022concept, jarvela2023human} and student agency analytics \cite{jaaskela2021student, hooshyar2023learning} to devise robust measures and mechanisms that capture learner agency \cite{fan2022towards}. This ensures that the essence of learning remains deeply rooted in human-centric principles.


\subsubsection{Automated Transcription}
\label{sec-data-transcription}

GenAI could also support multimodal LA solutions that relied on capturing and analysing learners' verbal communications \cite{chua2019technologies}. A major challenge exists in the accuracy and timeliness of transcribing learners' audio data as this process was often done in a post-hoc manner and mostly relied on manual efforts or external services, limiting the potential of these analytics in supporting real-time feedback and reflection practices \cite{yan2022scalability}. GenAI tools like Whisper and other speech-to-text models have demonstrated the potential to automate the transcription process \cite{rao2023transcribing}, opening up the opportunity to streamline the process of capturing and transcribing audio data to text data, which can be processed by established LA models or LLMs to extract meaningful insights, such as reflecting different teamwork constructs (e.g., situation awareness and close-loop communication) \cite{zhao2023mets}. While the research on the validity and reliability of GenAI in transcribing speech data still requires further empirical validations, as this line of research matures, it could potentially offer future LA research a less labour-intensive and integrated approach to cover meaningful insights within learners' verbal communication data.

\subsubsection{Dynamic Representations}
\label{sec-data-dr}

The diversity of learner data that may become available for future LA research may also be enriched by the prevalence of GenAI. Specifically, GenAI's capabilities in generating human-like multimodal content introduce novel opportunities to enrich the modality and interactivity of the data that can be used in LA research and practices, offering a more comprehensive understanding of the learner's learning process \cite{chiu2023impact, mazzoli2023enhancing, lee2023prompt}. Central to this potential is GenAI's capability to generate dynamic representations of learning \cite{zhang2011can}. Instead of the conventional reliance on static text or images, learners can now harness GenAI tools to transform their textual reflections or visual illustrations into animated sequences. For example, text-to-image models like Stable Diffusion, Midjourney, and Dall-E \cite{radford2021learning, chiu2023impact} can empower educators and learners to convert textual descriptions into detailed visual content. Such advancements could progress LA to diverse learning settings, such as art-focused STEAM education \cite{lee2023prompt}, where learners' continued interactions with these tools (e.g., prompts and iterations) may provide the data required to gain a deeper understanding of learners' cognitive processes like their proficiency in translating abstract concepts into tangible products \cite{fiorella2020creating, zhang2011can}. Likewise, the emergence of text-to-video generation tools, such as Runway Gen-2 and Make-A-Video \cite{singer2022make}, further enhance the richness of data available for analysis in LA. By transforming textual content into engaging video sequences, these tools can enhance the immersive nature of learning environments, shedding light on learners' reasoning and narrative abilities. Such dynamic representations are especially pivotal for capturing reliable data on learners with specific disabilities, like dyslexia \cite{long2007supporting}, reinforcing LA' commitment to inclusive education \cite{slee2018defining}. However, capturing learner agency during these novel learning processes is essential for ensuring the efficacy of LA.

\subsubsection{Synthetic Data}

Another opportunity that GenAI may contribute to the future of LA regarding data is supporting synthetic data initiatives. In LA research, synthetic data can serve as a valuable intermediary while waiting for actual data to become available, especially during bureaucratic processes like gaining approval from multiple ethics boards or navigating data ownership issues \cite{berg2016role}. Such synthetic data can offer researchers immediate access, especially those who may not have access to different data sources, enabling them to refine methodologies and evaluate the scalability and adaptivity of their solutions in data-rich environments. In a closely related field, educational data mining, synthetic data generation are already gaining momentum in data-mining contexts \cite{rahman2022educational}, suggesting that a similar approach can be extended to LA \cite{berg2016role}. For example, synthetic data can potentially support the development of privacy-preserving LA \cite{gursoy2016privacy}, provide LA training materials \cite{koester2015uom}, examine the inner workings of knowledge tracing models \cite{flanagan2022fine}, and contribute to the open-science practices in LA research \cite{dorodchi2019using}. While most of the previous studies aimed to generate synthetic data that mirrored the statistical attributes of learner data (e.g., demographic, educational data \cite{dorodchi2019using}), prior works on simulated learners illustrated its potential in playing the role of a tutoring training system for educators, learning partners for students, test fields for instructional designers to conduct formative evaluations with different learning systems \cite{flanagan2022fine, mccalla2013simulated}. However, these simulated learners were often limited in their ability to mimic actual human learners due to the complexity of human metacognition, cognition, and affection \cite{flanagan2022fine}. GenAI could potentially address these issues, especially with the recent emergence of AI agents, which are autonomous and adaptive AI that work towards a goal without continuous user input (e.g., AgentGPT \cite{rework2023agentgpt}). For example, these AI agents have also been used to simulate human behaviour at a scale of 25 agents, where they were able to engage in daily interaction, organise parties, and make new friends \cite{park2023generative}. Considering the versatility of these GenAI models, using these AI agents to play the role of learners, educators, and specialised roles, such as standard patients for medical students' diagnosis practice training \citep{song2022intelligent}, could open up a range of novel learning opportunities, and thus, generating rich data for future LA research to tackle. 

\subsection{Analytics}

The third step of the LA cycle involves extracting analytics from learner data to generate meaningful insights into the learning process. These analytics can be categorised into four different types \cite{solar2021learning}, including \textit{descriptive analytics} that use data aggregation and visualisation to understand historical trends and patterns of learning behaviours, \textit{diagnostic analytics} that use data mining and correlation analysis to understand the reasons behind observed trends and relationships, \textit{predictive analytics} that focus on combining historical data with statistical models to forecast potential trends and behaviours, and \textit{prescriptive analytics} that aim to recommend possible courses of action based on a combination of machine learning algorithms, business rules, and computational modelling. Apart from extracting these analytics from learner data, adequate reporting and presentation in LA dashboards are also essential to ensure their utility and impactfulness \cite{verbert2020learning, jivet2018license, matcha2019systematic}. Human-centred LA approaches are essential for designing effective LA that can be adopted, comprehended, and used successfully to support learning and teaching practices \cite{buckingham2019human, tsai_human-centered_2022}. We foresee four major potentials for GenAI to enhance the analysis, presentation, and informativeness of LA and diversify the interactions that educational stakeholders could potentially have with future LA dashboards. These potentials include supporting the analysis of unstructured data, enabling the multimodal illustration of LA, enriching the informativeness of LA with elaborated explanations, and contributing to the development of interactive LA. The following subsection elaborated on each of these potentials. 

\subsubsection{Unstructured Data Analysis}
\label{sec-data-unstrcutred}

Using natural language processing techniques to analyse and extract insights from unstructured text data is a rapidly evolving focus of LA research, spanning across educational discourse modelling \cite{dowell_modeling_2022} and writing analytics \cite{gibson_natural_2022, allen_natural_2022}. These data directly reflect the learners' experience, perceptions, and linguistic behaviours during learning tasks. They can provide different and potentially more enriched insights into their learning process and learning gains \cite{gray_practitioners_2022}. However, transforming unstructured text data to structured metrics that can be digested by LA systems to extract meaningful indicators is challenging. While conventional natural language processing techniques, such as Linguistic Inquiry and Word Count and Latent Dirichlet Allocation \cite{allen_natural_2022, xing2020identifying}, enabled opportunities to process unstructured text data, their utility is often limited to a specific task like extracting linguistic features or topic modelling, potentially limiting the practical values of the resulting insights. Prior studies using conventional natural language processing techniques for human-AI collaboration in qualitative analysis have identified limitations regarding the ability of these solutions to deal with the intricacies in the linguistic and semantic features of different language structures \cite{gebreegziabher2023patat, rietz2021cody}. On the contrary, GenAI, specifically LLMs, may offer more diversity and scalability in processing and analysing unstructured text data. For example, a study that used LLMs (GPT-3) for deductive coding found that with a pre-determined codebook, the model has achieved fair to substantial agreements with expert-coded results \cite{xiao2023supporting}. Furthermore, a recent study illustrated that state-of-the-art GenAI tools like ChatGPT demonstrated superior capabilities in capturing a variety of language structures. The ability of ChatGPT to elaborate on its decision was also deemed helpful to improve construct validity, pinpoint potential ambiguity in code definitions, and support human coders to achieve higher inter-rater reliability than prior tools (e.g., nCoder) \cite{zambrano2023ncoder}. This approach of using GenAI to explain its own decisions could potentially enhance the transparency of the decision-making process or at least provide additional information to cross-validate with humans. However, the validity and reliability of GenAI in analysing unstructured text data still require further investigation.

\subsubsection{Multimodal Illustration}
\label{sec-data-mi}

As discussed in Section \ref{sec-data-dr}, GenAI can potentially enhance the dynamic representations of learning resources and learner data. Similar opportunities may also apply to the illustration and presentation of LA. Prior LA research has demonstrated the benefits of data-driven personalised feedback in improving learners' academic achievement \cite{pardo2019using, lim2021changes}. One of these systems, OnTask \cite{pardo2018ontask}, offers an interface that allows educators to transform learner data into specific rules for email generation, thereby enabling personalised feedback for each learner and ensuring that the feedback is directly relevant to their learning progress. For such personalised feedback, GenAI offers the opportunity to move beyond mere text or graphical feedback but also enables the generation of audio and potential video feedback. For example, using 3D diffusion model \cite{wang2023rodin} and text-to-speech models (e.g., Meta's Voicebox \cite{le2023voicebox}), educators can generate a digital representation of themselves and deliver the feedback content through narrated voice instead of merely text. Apart from benefiting learners with dyslexia \cite{long2007supporting} or preferring multimedia learning \cite{mayer2005cognitive}, there is also some evidence showing that audio and video feedback was perceived as more personal, fostering learner comprehension, and more dynamic and engaging than written feedback \cite{mccarthy2015evaluating, orlando2016comparison}. Yet, the lack of usage of these feedback modalities was mainly associated with the workload and technical challenges in creation and distribution \cite{mccarthy2015evaluating}, which can be potentially addressed with the advancement in GenAI and the emergence of low-tech solutions. However, a cautious approach must be taken when using these 'deepfake' technologies to generate personable content as they present unique ethical and cybersecurity challenges (e.g., identity theft and fake information) \cite{westerlund2019emergence}.

\subsubsection{Explanatory Analytics}

Besides supporting dynamic representations, GenAI can also augment existing LA visualisations by providing elaborated explanations. This potential resonates with the emerging studies that utilised data storytelling to enhance learners' and educators' interpretation of LA \cite{martinez2020data, nieto2022beyond}. For example, prior studies have illustrated that adding text narrative and a layered approach to LA visualisations has provoked deeper reflection in educators \cite{martinez2020data} and supports the sensemaking of complex visualisations \cite{nieto2022beyond}. Moving away from exploratory and towards \textit{explanatory visual LA} could facilitate learners' and educators' ease of interpretation and provide more actionable insights \cite{echeverria2018exploratory}. However, a current limitation in this line of research is the dependency on structured learning activity, which is required to formulate the rules and protocols that generate the explanations. This limitation undermines the utility of explanatory analytics in open-ended learning tasks where learning behaviours were less predictable \cite{martinez2020data}. GenAI could potentially contribute to addressing such limitations. One potential approach is to use GenAI (e.g., AgentGPT \cite{rework2023agentgpt}) to create simulated learners and run multiple learning simulations, in which clustering techniques and reinforcement learning can be applied to identify prevalent learning strategies \cite{flanagan2022fine, mccalla2013simulated}. These simulated learning strategies can derive rules and protocols for supporting explanatory analytics in open-ended learning tasks. Another potentially more generalisable approach is to use retrieval-augmented generation \cite{lewis2020retrieval} with LLMs to generate more accurate and context-sensitive explanations. This approach could potentially be achievable by storing previous knowledge of learning tasks (e.g., learning design and theoretical underpinnings) in an embedding database \cite{ham2020end}, where LLMs can draw on for more contextual information when generating explanations for visual LA (visualisations can be generated automatically by LLMs with a code interpreter plugin like ChatGPT-4). 

\subsubsection{Interactive Analytics}
\label{sec-analytics-ia}
GenAI can also help to address several essential issues around LA dashboards, including one-size-does-not-fit-all, actionability, and data literacy. To address these issues, Verbert et al. \cite{verbert2020learning} emphasised the need to move beyond building LA dashboards that serve merely as a one-way information route, where users receive information and potentially get more explanations through interaction (e.g., click on a data point). GenAI could transform these conventional dashboards into dynamic, conversational systems, where LA dashboards become interactive platforms that foster the collaboration between the user and the dashboard in achieving a shared goal like enhancing learning outcomes \cite{charleer2017learning}. For example, using retrieval-augmented generation \cite{lewis2020retrieval} to connect GenAI with course resources and relevant theoretical frameworks (e.g., self-regulated learning \cite{winne_learning_2022, matcha2019systematic}) could support AI-powered dashboards to proactively recommend resources, interventions, or learning strategies to learners based on their current performance and trajectory instead of merely presenting static data. By infusing relevant contextual and theoretical knowledge into AI-powered dashboards, LA dashboards can also become more interactive, where users can engage in conversation with the dashboards (e.g., a chatbot for the dashboard) to enhance their interpretation and internalisation of the analytics (e.g., provoking self-reflection and planning behaviours \cite{jivet2017awareness}). However, ensuring that AI-powered dashboards respect data privacy (e.g., transparency in data usage) and avoid potential inaccuracy and biases in their communication and recommendations is crucial \cite{verbert2020learning}. 

\subsection{Interventions}

The final step to complete the LA cycle is going beyond methodologies, instruments, and scientific results and applying LA in practice to deliver interventions that optimise the learning process and outcomes \cite{clow2012learning}. Such interventions could be providing adequate support to students who have been identified as at-risk through predictive LA \cite{herodotou2019empowering} or utilising multimodal LA to reflect and improve learners' oral presentation skills \cite{ochoa2018rap}. While this final step is essential to make LA more impactful and contribute to addressing immediate and long-standing challenges in education \cite{gavsevic2015let, wise2021makes}, it is also the step that is most underdeveloped in existing LA literature \cite{papamitsiou2020childhood} with merely a small amount of studies that demonstrated practical benefits on learning outcomes, supports, and teaching \cite{viberg2018current}. Regarding supporting LA intervention initiatives, we foresee three major opportunities that GenAI may offer to promote such interventions' development, delivery, and evaluation. These opportunities include enhancing the level of personalisation available through LA interventions, improving the adaptivity of these interventions from a longitudinal perspective, and broadening the accessibility of LA across different learners and regions of the world. The following sections elaborate on these opportunities and discuss the associated challenges.

\subsubsection{Personalisation} 

Prior studies on learners' attitudes and expectations of LA have consistently identified personalisation as one of the expected values of LA \cite{pardo2019using, schumacher2018features, roberts2017give}. For example, learners desired LA dashboards that generate personalised feedback and recommendations to support further improvements \cite{ochoa2018rap, karaoglan2020student}. However, personalisation is a deeply human-centred approach, where different learners may have diverse perspectives. For example, some learners may feel uncomfortable and unfair for others to receive additional learning content or support as they could potentially be disadvantaged during the process of personalisation, others may prefer automated and unidentifiable notifications over personalised and identifiable emails from educators \cite{roberts2017give, pardo2018ontask}. Likewise, personalisation is also associated with dynamic representations (Section \ref{sec-data-dr}) and multimodal illustration (Section \ref{sec-data-mi}), as different learners may have different preferences over the reporting and presentation of LA \cite{mccarthy2015evaluating, orlando2016comparison}. These differences in individual preferences pose challenges in developing LA interventions that could cater to the requirements of different learner groups. To achieve such in-depth levels of personalisation, LA interventions should be developed based on the needs, knowledge, performances, learning process and contexts of individual learners \cite{buckingham2019human,roberts2017give}. While this level of personalisation seems almost impossible to achieve, we foresee the potential to move towards this goal by utilising the aforementioned capabilities of GenAI. For example, the ability of GenAI to transform unstructured text into structured data \cite{xiao2023supporting}, analyse diverse linguistic structures \cite{zambrano2023ncoder}, and generate coherent feedback \cite{dai2023can} can be integrated into functions that foster the reciprocal communications between learners and LA interventions \cite{verbert2020learning}, where these communications (e.g. chat history) can be processed to update learners' current knowledge, performances, and learning process and contexts and informing researchers and practices about any emerging needs in a real-time manner (e.g., automatically summarised report). Additionally, combined with GenAI's ability to generate multimodal illustrations, learners could control what, where, and how they would like to receive LA interventions. However, this take on personalisation falls into the philosophical debate of individual autonomy \cite{hannafin1997foundations} versus paternalism \cite{dworkin2014paternalism}, which demands LA researchers and practitioners to consider the balance between structured guidance and learner autonomy over personalisation for different learner groups (e.g., age and culture).

\subsubsection{Adaptive Intervention} 

Another opportunity that GenAI may bring to the delivery of LA involves its capabilities to support adaptive intervention. Effective LA and technologies should provide learners with adaptive support tailored to their current knowledge, performances, learning process and contexts \cite{han2021learning, forsyth2016maximizing}. Prior LA research has focused on creating learner profiles that capture the essence of a learner's history, preferences, and performance metrics to tailor educational experiences \cite{corrin2017using, zamecnik2022exploring, barthakur2023advancing}. These profiles, based on demographics and accumulated trace data over time, have been instrumental in informing researchers and educators about learners' learning behaviours (e.g., help-seeking \cite{corrin2017using}) and providing adequate support to optimise their educational experiences \cite{barthakur2023advancing}. We foresee an exciting opportunity for integrating GenAI into prior LA research on learner profiles \cite{corrin2017using, zamecnik2022exploring, barthakur2023advancing}. GenAI's ability to analyse vast and diverse datasets (as discussed in Section \ref{sec-data-unstrcutred}) allows it to construct detailed learner profiles, taking into account various competencies, attributes, and experiences (e.g., captured through interactive analytics; Section \ref{sec-analytics-ia}). Unlike conventional rule-based or static models, which may often be confined to specific disciplinary domains, GenAI can integrate data across multiple domains, offering a more holistic view of a learner's learning trajectory. Additionally, by connecting GenAI with external knowledge (through multiple embedding databases), it could potentially navigate the complexities of various educational spaces (e.g., technical, social, and political \cite{barthakur2023advancing}), offering insights that are not just confined to individual contexts but span across multiple educational scenarios. This comprehensive profiling can aid in bridging the gap between current assessment practices and a more adaptive, learner-centred approach \cite{hannafin1997foundations}. The potential of GenAI also lies in its ability to be integrated with previously validated methodologies and instruments. For example, recent system infrastructural progress of the FLoRA engine (a LA platform for supporting self-regulated learning) \cite{li2023flora} has combined learners' prior knowledge and self-regulated learning process to generate adaptive prompts for ChatGPT to provide learners with adaptive scaffolding during their learning process. Such integration could advance future LA interventions, making them more adaptive and holistic. 

\subsubsection{Accessibility} 

Regarding accessibility, GenAI presents a nuanced scenario for future LA interventions. While its capabilities in dynamic representations could indeed be beneficial for learners with specific challenges or disabilities, such as dyslexia \cite{long2007supporting}, there is an imminent concern that the economic implications might overshadow these advantages \cite{van2023chatgpt}. For those individuals and institutions with financial capacities, leveraging commercial GenAI tools or deploying open-source versions can potentially enhance the depth, interactivity, and multimodality of their LA interventions, as elaborated in Sections \ref{sec-data-mi} and \ref{sec-analytics-ia}. However, the computational resources associated with the development, deployment, and operation of GenAI are concerning, potentially mirroring or even exceeding their capabilities \cite{van2023chatgpt, kasneci2023chatgpt}. Such economic barriers could widen the digital divide, placing advanced GenAI-powered LA tools out of reach for many learners, especially those in under-resourced settings \cite{pontual2020applications}. Likewise, while GenAI could potentially support personalised learning pathways, it can amplify the looming risk of disadvantaging large numbers of people and creating a two-tiered education system: one where affluent learners enjoy personalised, dynamic, and interactive learning experiences, and another where less privileged learners are stuck with static, one-size-fits-all content \cite{selwyn2019s}. The ethical implications of this discrepancy are significant. It raises questions about equity in education and the potential for GenAI and LA to heighten existing inequalities rather than alleviate them \cite{yan2022scalability, slee2018defining}. Additionally, while the initial investment in GenAI might be high, there is also the ongoing cost of maintenance, updates, and ensuring compatibility with educational methodologies and policies \cite{tsai2017learning}. As LA advances, it is crucial to consider the pedagogical implications of GenAI and its broader socioeconomic impact. A balanced approach, where the benefits of GenAI are made available more equitably while mitigating its challenges, will be essential for the future of inclusive and effective LA interventions \cite{slee2018defining}. 
\section{Future Research Endeavours with Generative AI}

We summarised the contextualised opportunities and challenges of GenAI through the LA cycle \cite{clow2012learning} into six potential research endeavours for future studies to explore and investigate the synergy between LA and GenAI. Through these endeavours, the LA community can effectively and ethically leverage the technological advancement of GenAI and fulfil its foundational purpose of optimising learning for all learners.

\noindent\textbf{Fist Endeavour: Hybrid Human-AI Learning Paradigms.} From the learner aspect, as the lines blur between learners and GenAI tools, future LA research should delve deep into frameworks and methodologies that advocate for a collaborative learning experience between humans and AI. Investigating these hybrid models will offer in-depth insights into learners and their learning process in the age of GenAI \cite{jarvela2023human, jarvela2023hybrid, zhang2021ideal, national2021human}. 

\noindent\textbf{Second Endeavour: Advanced Analysis of Unstructured Learning Data.} From the data aspect, given the richness of unstructured text data in reflecting learners' cognitive processes, future LA research should focus on leveraging GenAI for a nuanced analysis of such data. This would involve processing discourse, reflections, and other forms of textual data to gain deeper insights into the learning process \cite{dowell_modeling_2022, gibson_natural_2022, xiao2023supporting, zambrano2023ncoder}. 

\noindent\textbf{Third Endeavour: Enhanced Visual and Explanatory Analytics.} From the analytics aspect, GenAI can augment conventional LA dashboards and visualisations, offering more detailed, contextual explanations and data storytelling. Future research should look at integrating GenAI to make LA more actionable, interactive, and user-friendly, ensuring a deeper understanding and reflection for both learners and educators. \cite{martinez2020data, nieto2022beyond, charleer2017learning, jivet2017awareness}. 

\noindent\textbf{Forth Endeavour: Deep Personalisation and Adaptive Interventions.} From the intervention aspect, with GenAI's capability to offer real-time updates and create detailed learner profiles, future LA interventions can achieve deeper personalisation. This requires research into how GenAI can be integrated to ensure interventions remain relevant and effective throughout a learner's educational journey \cite{buckingham2019human, roberts2017give, han2021learning, barthakur2023advancing}.

\noindent\textbf{Fifth Endeavour: Enhancing Accessibility in LA with GenAI.} From the intervention aspect, future research should ensure that advancements in GenAI-powered interventions are accessible to all learners, regardless of their socio-economic background. This requires addressing economic barriers and ensuring that LA interventions remain inclusive and equitable \cite{long2007supporting, slee2018defining, yan2022scalability}.

\noindent\textbf{Sixth Endeavour: Ethical and Equitable Integration of GenAI in LA.} Overall, as GenAI becomes more embedded in LA, addressing ethical concerns becomes essential. Future research should focus on ensuring the integration of GenAI in LA respects data privacy, maintains ethical standards, and does not amplify existing educational inequalities and problematic practices, such as gender bias and discrimination \cite{tsai2020privacy, slee2018defining, yan2022scalability, verbert2020learning}.
\section{Final Remark}

For LA to achieve its promise in optimising learning, the LA community needs to venture beyond merely uncovering insights about learning or developing innovative methodologies and analytics. The promise of LA lies in its actionable application---addressing real-world educational problems and benefiting both teaching and learning processes \cite{mazzoli2023enhancing,wise2021makes, papamitsiou2020childhood}. This position piece illustrates a promising synergy between GenAI and LA. Such a combination has the potential to support systemic changes in education, making LA more comprehensive, evidence-based, and aligned with the real-world complexities of education \cite{tsai2017learning}. Yet, despite this optimistic vision, we must be vigilant. GenAI also introduces new theoretical complications, such as redefining the very essence of a 'learner', and has boarder socioeconomic impacts, notably the risk of widening the digital divide. As the field of LA moves forward, these considerations must be at the forefront of a collective endeavour among LA researchers, practitioners, and policymakers, ensuring an inclusive future for both LA and education in the era of GenAI.
\begin{acks}
This research was at least in part funded by the Australian Research Council (DP210100060) and Jacobs Foundation (Research Fellowship).
\end{acks}

\bibliographystyle{ACM-Reference-Format}
\bibliography{0_reference.bib}






\end{document}